\documentclass{aastex63}

\graphicspath{{./}{figures/}}

\received{March 30, 2020}
\revised{ ?, 2020}
\accepted{?, 2020}
\submitjournal{PSJ}

\shorttitle{km-scale Jovian Irregulars}

\begin{document}

\title{The population of km-scale retrograde jovian irregular moons}

\correspondingauthor{Edward Ashton}
\email{eashton@phas.ubc.ca, gladman@astro.ubc.ca}

\author{Edward Ashton}
\affil{University of British Columbia \\
Dept. of Physics and Astronomy \\ 
6224 Agricultural Road \\
Vancouver, BC, V6T 1Z1 Canada}

\author{Matthew Beaudoin}
\affil{University of British Columbia \\
Dept. of Physics and Astronomy \\ 
6224 Agricultural Road \\
Vancouver, BC, V6T 1Z1 Canada}

\author[0000-0002-0283-2260]{Brett Gladman}
\affil{University of British Columbia \\
Dept. of Physics and Astronomy \\ 
6224 Agricultural Road \\
Vancouver, BC, V6T 1Z1 Canada}



\begin{abstract}

We have searched a 2010 archival data set from the Canada-France-Hawaii Telescope for very small (km-scale) irregular moons of Jupiter in order to constrain the size distribution of these moons down to radii of $\sim 400$~m, discovering 52 objects which are moving with Jupiter-like on-sky rates and are nearly certainly irregular moons. The four brightest detections, and seven in total, were all then linked to known jovian moons.
Extrapolating our characterized detections (those down to magnitude $m_r=25.7$) to the entire retrograde circum-jovian population, we estimate the population of radius $>0.4$~km moons to be 600 (within a factor of 2).  At the faintest magnitudes we find a relatively shallow luminosity function of 
exponential index $\alpha= 0.29 \pm 0.15$, corresponding to a differential diameter power law of index $q\simeq 2.5$.

\end{abstract}



\section{Introduction} \label{sec:intro}

Irregular moons were likely once Sun-orbiting minor bodies that were captured by a giant planet early on in the Solar System's history. The mechanism that changed the object from a heliocentric to a planetocentric orbit is still uncertain, although multiple theories have been suggested: gas drag, pull down due to sudden mass growth and three-body interactions\citep{Nicholson}.
Short orbital periods, eccentric orbits and the relatively small volume of space that irregular moons occupy results in collisions significantly altering the initial population of irregular moons into today's size distribution. Still, the current size distribution of irregular moons provides some constraint on dynamical models of irregular moon systems and their initial population. 

Only nine irregular moons of Jupiter were known before the year 1999. Thanks to wide-field Charge Coupled Device (CCD) cameras, an explosion of new discoveries occurred around the turn of the millennium. Since then, the number of new discoveries has dropped off, with the only survey of note in the last 10 years being a 2017 study finding 12 new jovian irregular moons \citep{Sheppard2018}. Before we started our work there were 71 known jovian irregular moons, 10 of them with direct orbits and 61 with retrograde.

Irregular moons, like other small body populations, have size distributions that appear to obey an exponential law: $N(<H) \propto 10^{\alpha H}$, where $H$ is the absolute magnitude of the moon, $N(<H)$ is the number of moons that have a $H$ magnitude of ’H’ or less and $\alpha$ is the logarithmic slope. Since moons around a single giant planet are roughly the same distance from Earth, $H$ can be replaced with an apparent magnitude, $m_r$. The size distribution of jovians is shallow at large sizes with $\alpha \approx 0.2$ for $m_R < 19~(r > 10~\text{km})$\footnote{Approximate sizes are calculated using a 4\% albedo.} and $\alpha > 0.5$ for $m_R > 21.5~(r < 4~\text{km})$ \citep{Nicholson,SheppardnJewitt}. In the intermediate magnitude range the size distribution exhibits `strong flattening' ($\alpha < 0.2$) \citep{SheppardnJewitt}. 

In this paper we describe the data set and our method for finding jovian moons (Section \ref{sec:Methods}) and how we produced our size distribution and subsequent analysis (Section \ref{sec:Analysis}).

\section{Current Data Set and Reduction methods}
\label{sec:Methods}

We analyzed a $\sim$3-hour archival Canada France Hawaii Telescope (CFHT) data set from 2010 that was originally taken to recover (successfully) a known jovian irregular moon; in addition, two new moons (Jup LI and LII) above the single exposure limit of the frames were discovered and reported \citep{Alexandersen2012}.
Although it was clear that this data set could have been shifted and stacked to reveal fainter moons, 
substantial ranges of on-sky rates and angles would need to be searched to find unknown jovian irregulars, so this process was not attempted at the time.
In summer 2019 we decided to cover this large shift-and-stack parameter space for the data set, as the nominal expected depth of $m_r\simeq25.5$ exceeded the depth of previous searches by at least a magnitude.
This made a cumulative luminosity function study down to 
diameters below 1 km accessible, even if tracking the 
objects to determine orbits was impossible.

The Sept 8/2010 UT field consisted of a single one square-degree CFHT MegaPrime field of 36 CCDs, centered at an offset of $1.5^{\circ}$ West and $0.04^{\circ}$ North of Jupiter.
The images were unbinned, with a 0.186 \arcsec \ per pixel scale.
There were 60 sequential 140-second exposures (with a 40-second CCD readout time) which meant this sequence lasted 3 hours.
With a jovian on-sky motion of 19.2 \arcsec/hr, longer exposures were precluded.  
The single exposures were sufficient to recover the previously-known moons down to magnitude $\simeq24$ in the best-seeing frames, and thus with 60 exposures we expect to be able to see moons down to about 26th magnitude at the very limit of a stacked sequence.

The order of image processing was as follows: all images were aligned to the first image of the sequence; artificial moving objects were implanted in each image (details in next paragraph); the images were flux scaled relative to reference stars in the first image, and a $25\times25$-pixel sized boxcar filter was then applied to background subtract the images (for details on these two processes see \citet{Gladman2001}).  This method is quite effective at minimizing stellar confusion as the moons move in front of them over the exposure sequence.

To determine our efficiency at finding moons, a random number of 600--650 artificial moons were implanted into each CCD. The magnitude, on-sky rate and position angle (PA)\footnote{Here position angle is the direction of motion of an object measured anti-clockwise from direct north.} of the implanted moons were drawn from a uniform distribution ranging from 24--26.2, 83--127 pix/hr and 238--252$^{\circ}$ respectively. The rate and PA ranges were chosen to go slightly beyond the minimum and maximum values for all known jovian irregulars excluding Themisto (PA = 232.2$^{\circ}$). We believe our exclusion of Themisto is justifiable due to it travelling close to Jupiter at the time (only 13' away), and is thus not reflective of the motion of a moon in our field, which is much further away from the planet. The combination of image quality and rates of the implanted moons means the implanted moons should be trailed. The implanting software simulates trailing by splitting the signal into ten equal pieces and implant each piece, with equal time space, in each exposure. We only implanted moons on parts of the CCDs where we knew that the fastest moon would not move off the CCD during the 3-hour sequence, as this was also how the frames would
be later trimmed after shifting.

Once this processing was complete, the image set was shifted at a grid of different rates and PAs, and then combined using the median value at each pixel. To remove any cosmic rays or bad pixels and to lessen the presence of stars, we rejected the five highest values along with the lowest value for each pixel while combining the images. The range of shift rates and PAs were chosen to be slightly smaller than the implanted range; 85 to 125 pix/hr and 240--250$^{\circ}$ respectively. Using step sizes of 2 pix/hr and 2$^{\circ}$ produced 6 different shift PAs and 21 different shift rates, with a total of $6 \times 21 = 126$ different recombinations. 
We trimmed away from the stacked images any stacked pixel which would have resulted in a moon starting on that pixel leaving the field over
the 3-hour sequence.
This procedure, along with the original (smaller) CCD gaps,
meant we were able to search about $80\%$ of the one-degree outer boundary (and all of the retained coverage reached our full magnitude depth).
There are thus 36 adjacent 'mini-fields' in our search.

All rates and PAs were searched methodically by two human operators using a five-rate blinking sequence, with the fastest rate of the last five rates becoming the slowest rate of the next sequence (to provide overlap). By blinking multiple rates at a time, moons can be easily identified by their characteristic pattern of coming in and out of `focus' as the recombination rate and PA gets closer and further from the moons actual rate and PA. Initially, each CCD was searched by both operators. After searching 3 CCDs, the detection of the two operators were deemed similar enough that only one operator searched subsequent single CCDs, in order to save time.

Even if it may be possible to somewhat improve this process in various ways, we note that due to
the calibrated implantation of artificial objects, the effectiveness of our search {\it is}
accurately measured and the subsequent debiasing is correct.  That is, perhaps it is possible to search the data set more deeply, but {\it this} search's effectiveness has been determined.

\section{Results}
\label{sec:Analysis}

\subsection{Detection efficiency}

After all rates and PAs were searched over all CCDs, the objects we detected were compared with the implanted moons. Any implanted object that was matched (within a tight tolerance) to a detection was labeled `found'. Any object found that was unable to be matched to an implanted object became a candidate moon.

\begin{figure}[ht]
\plotone{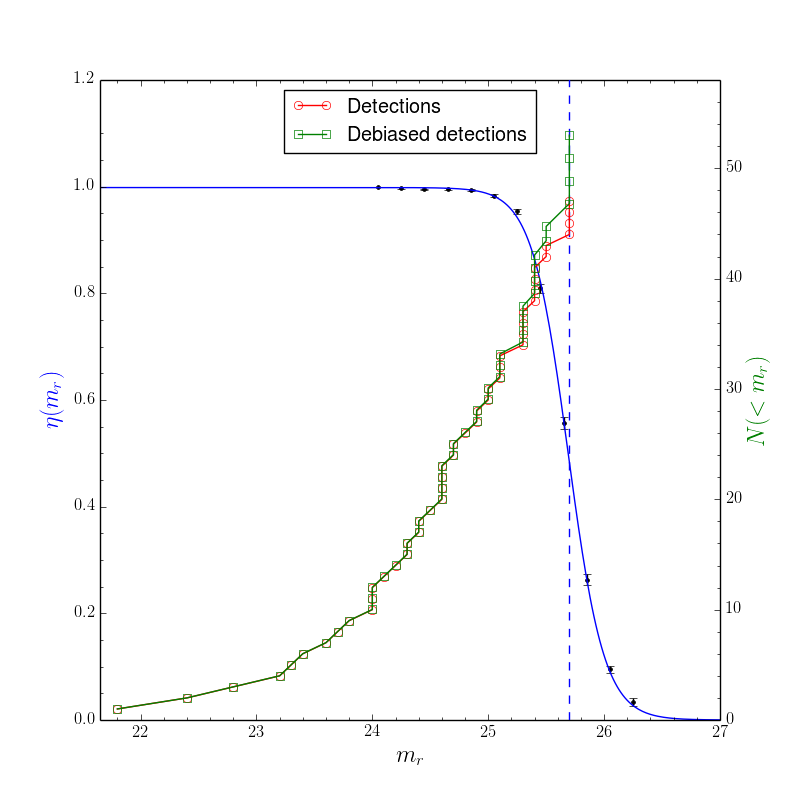}
\caption{This project's magnitude detection efficiency and cumulative luminosity function. Black data points show the binned fraction of detected implanted objects over the whole field and the best fit hyperbolic tangent function (blue solid curve), which we use as our detection efficiency function $\eta(m_r)$. The characterisation limit (blue dashed line) was chosen to be where our detection efficiency dropped to 0.5. The red line represents the cumulative number of our characterised detections. Our debiased number of detections in the field (green line) were calculated by weighting each detection by 1/$\eta(m_r)$, where $m_r$ is the detected objects r-band magnitude.}
\label{fig:effifunc}
\end{figure}

The fraction of implanted objects found, as a function of magnitude, over the whole field is very close to $1$ on the bright end. 
Despite a typical moving source's path crossing several background galaxies, stellar halos, and/or bad-pixel columns, essentially all sources brighter than
25th magnitude were easily recovered.
The detection efficiency then starts to drop around $m_r = 25$ and falls to 0.5 by 
$m_r = 25.7$ (see Fig. \ref{fig:effifunc}, black points and blue curve). 
We fit this fraction with a hyperbolic tangent function $\eta(m_r) = \frac{A}{2}(1-\tanh(\frac{m_r-\mu}{\delta}))$, where $A\simeq1$ is the fraction of bright objects that are detected, $\mu$ is the magnitude where the fraction drops to $A/2\simeq0.5$, and $\delta$ is a `width' of the drop. The curve of best fit for the implanted moons (blue line in Fig. \ref{fig:effifunc}) has the parameters $A = 0.998, \mu = 25.69, \delta = 0.31$, which we use as our detection efficiency. For the luminosity function study we term the `characterisation limit' to be where the detection efficiency drops to $0.5$. Therefore our characterisation limit is $m_r = \mu = 25.7$; brighter than this limit we have  
confidence that we can accurately debias the detected sample. Performing the same fit on various sub-regions of our field enabled the detection of a tiny drop of about 0.1 mags in the 50\% limit going from the side of the field that is furthest from Jupiter to the closest. Thus there is an uncertainty of 0.05 mags on our characterisation limit.

The fraction of implanted moons that were found as a function of rate and PA is mostly constant except for small drop offs (about $10\%$) at the extreme minimum and maximum implanted values. This drop off is due to the implanted ranges being slightly larger than the search ranges. Since only a small fraction of our moon candidates have rates and/or PAs outside our search ranges, any effects caused by these drop offs will be negligible.

\subsection{Detections}

\begin{figure}[ht]
\plotone{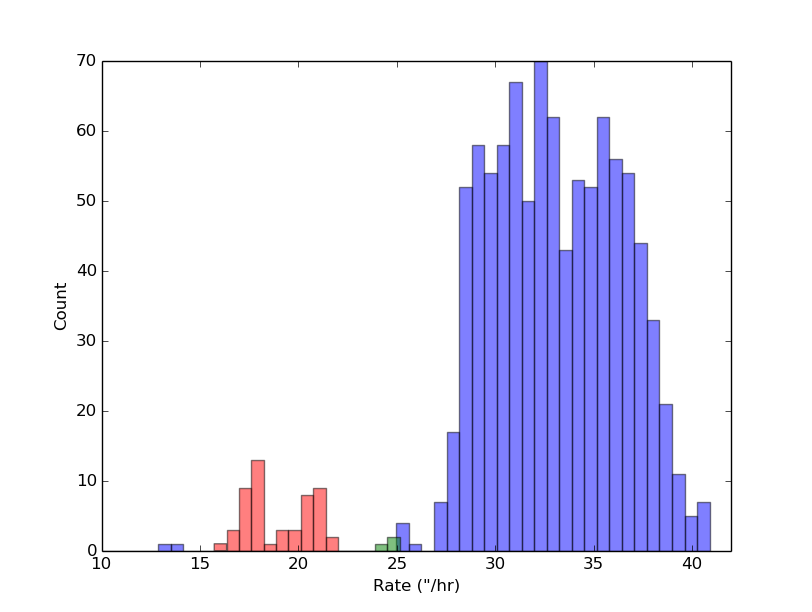}
\caption{A histogram of the on-sky rate our characterised detections (red) compared with known minor bodies that are within 3$^{\circ}$ of the centre of our field and have position angles
for their velocity vectors between 238$^{\circ}$ and 252$^{\circ}$ at the start of the observing sequence (blue). The green histogram bins show the three detections that are traveling beyond the jovian irregular rates (Table~\ref{tab:uncharacdet}), which we thus believe are asteroids on the tail of the asteroid rate distribution. Known minor bodies that are not shown are Trans-Neptunian Objects with rates less than 5 \arcsec/hr and Near Earth Objects with a rate of above 50 \arcsec/hr.}
\label{fig:ratehist}
\end{figure}

During our search activities, 55 moon candidates were discovered. 
Three of these candidates were travelling faster than the fastest rate search, which raised doubt as to whether they are jovian moons. 
Examining the distribution of on-sky rates of our detections, there is a clear separation between these three fast moving objects (green histogram in 
Fig.~\ref{fig:ratehist}) and the rest of the detections (red histogram)\footnote{The double peak in the rate distribution of the detected objects is centred on the jovian rate and believed to be due to the majority of the known retrograde moons having projected orbits that extend beyond our field. As such, very few come to rest relative to Jupiter in our field. Thus almost all of our detections should be going faster or slower than Jupiter's rate than at it.}. We compared our detections with all minor bodies in the IAU Minor Planet database that are within three degrees of our field centre at the time of observation and have PAs between 238$^{\circ}$ and 252$^{\circ}$ (a range that encompasses the known moons). The three fast moving detections are right on the tail of the asteroid distribution (25--41 \arcsec/hr), which leads us to believe these objects are in fact slow-moving asteroids, and will thus not be included in our analysis of the jovian luminosity function. The region in which our moon candidates lie, 15--23 \arcsec/hr, contain no known minor planet, one line of evidence that strongly indicates our candidates are indeed jovian moons.

Four of the candidates were beyond the characterisation limit of $m_r = 25.7$ and another one of the candidate moon was found outside our search area by chance while performing astrometry and photometry on a different object\footnote{This extra detection (j10r113a22) had moved into a chip gap during the observation sequence, thus outside of our characterised search area, but it was bright enough to be seen on a subset of the images.}. The three fast, four faint and one extra candidates are not included in the characterised sample, which thus consists of 47 moons. A full list of the characterised and uncharacterised detections are found in Tables \ref{tab:characdet} $\&$ \ref{tab:uncharacdet} respectively. The cumulative number of our detections as a function of magnitude is shown in Fig. \ref{fig:effifunc} (red line). To debias the detections, we weighted them by the detection efficiency. Each detection became $1/\eta$ detections, where $\eta$ is the detection efficiency at the object's measured magnitude. Only near the characterisation limit does the cumulative number of debiased detections (green line) becomes noticeably different from the unbiased detections, due to $\eta\simeq1$ except at the sharp drop off near the limit; because we have chosen to only debias down to the 50\% detection
efficiency, our population estimate is little affected by the efficiency
correction.

We provide the astrometric entries (in standard Minor Planet Center format) in Table 3.

\setcounter{table}{0}
\begin{table}[ht]
\renewcommand{\thetable}{\arabic{table}}
\centering
\caption{A list of all characterised detections. We give our internal designation, the MPC designation if it is previously known, and the moon's $r$-band magnitude. Our designations start with a `j' (for Jupiter) followed by the CCD number the moon was found on, then an `r' (for rate), followed by the rate (in pixels/hr) which gave the best recombination, then an `a' (for angle), and lastly the angle (in degrees) which gave the best recombination.Note: the angle we use here is NOT the PA but 270$^{\circ}$ minus the PA.
Known designations in ( ) indicate that the identification was not simply at the nominal position based on the IAU Minor Planet Center ephemeris.} \label{tab:characdet}
\begin{tabular}{ccc}
\tablewidth{0pt}
\hline
\hline
\colhead{Our designation} & \colhead{Known designation} & \colhead{$m_r$} \\
\hline
\decimals
 j20r97a28  & Hermippe     & 21.8 \\
 j32r104a24    & Erinome      &          22.4 \\
 j31r105a28   &   (S/2003 J16)  &     22.8  \\
 j09r89a20 & Jup LIX       &          23.2  \\
 j35r115a26   & -  &       23.3 \\
 j30r97a20    & -  &       23.4 \\
 j25r108a26 & Jup LII   &          23.6 \\
 j22r91a19    & - &        23.7 \\
 j31r113a26   & (Jup LXIX) &        23.8 \\
 j11r99a26    & - &      24.0 \\
 j22r94a24    & -   &      24.0 \\
 j27r97a26    & -  &       24.0 \\
 j03r94a24    & -  &       24.1 \\
 j23r95a24 & Jup LI        &          24.2 \\
 j20r112a25   & - &        24.3 \\
 j32r98a26    & - &        24.3 \\
 j23r113a27   & - &        24.4 \\
 j30r114a28   & - &        24.4 \\
 j24r97a22    & - &        24.5 \\
 j00r92a25    & - &        24.6 \\
 j22r98a18    & - &        24.6 \\
 j27r118a26   & - &        24.6 \\
 j33r98a22    & - &        24.6 \\
 j16r108a21   & - &        24.7 \\
 j20r109a31   & - &        24.7 \\
 j24r98a22    & - &        24.8 \\
 j29r112a25   & - &   24.9 \\
 j31r94a26    & - &     24.9 \\
 j21r97a26    & - &    25.0 \\
 j32r97a26    & - &     25.0 \\
 j13r105a20   & -    &     25.1 \\
 j24r113a22   & -    &     25.1 \\
 j28r95a26    & -    &     25.1 \\
 j20r93a28    & -    &     25.3 \\
 j22r110a23   & -    &     25.3 \\
 j27r116a25   & -    &     25.3 \\
 j31r97a22    & -    &     25.3 \\
 j18r92a22    & -    &     25.4 \\
 j30r114a19   & -    &     25.4 \\
 j32r98a24    & -    &     25.4 \\
 j34r96a22    & -    &     25.4 \\
 j09r109a21   & -    &     25.5 \\
 j33r107a23   & -    &     25.5 \\
 j09r111a20   & -    &     25.7 \\
 j09r93a20    & -    &     25.7 \\
 j10r109a22   & -    &     25.7 \\
 j12r113a25   & -    &     25.7 \\
\hline
\end{tabular}
\end{table}

\setcounter{table}{1}
\begin{table}[ht]
\renewcommand{\thetable}{\arabic{table}}
\centering
\caption{A list of all of our uncharacterised detections with our internal designation and the moon's magnitude. The naming convention used is same as what is described in the caption of Table \ref{tab:characdet}, except the three objects that we believe are moving too fast to be moon 
(see Fig.~\ref{fig:ratehist})
start with an `f' instead of a `j'. 
\label{tab:uncharacdet} }
\begin{tabular}{ccc}
\tablewidth{0pt}
\hline
\hline
\colhead{Our designation} &  \colhead{$m_r$} & \colhead{Reason for being uncharacterised}\\
\hline
\decimals
 j10r113a22 & 24.0 & Found outside search area \\
 f11r133a19   &      22.9 & Moving too fast  \\
 f02r131a22  &       23.3 & Moving too fast   \\
 f17r136a22   &  24.9 & Moving too fast  \\
 j12r115a26    &    25.9 & Beyond mag limit  \\
 j07r112a24    &     26.2 & Beyond mag limit \\
 j23r114a20   &   26.4 & Beyond mag limit \\
 j10r89a30     &    26.5 & Beyond mag limit \\
\hline
\end{tabular}
\end{table}

\subsection{Serendipitous Tracking}
\label{sec:tracking}

It was obvious that most of the irregular moon detections found in this pencil-beam search would be beyond the magnitude limit of other available data (because it is only another shift-and-stack pencil-beam study that could to descend beyond 24th magnitude for these rapidly-moving targets).
Nevertheless, we were able to identify some additional observations beyond the night of
detection.
We believe that essentially every object in the 15--23 \arcsec/hr rate range is a jovian moon.

Although there are hundreds of bright (visible on single images) asteroids in this field,
the only two objects with $m_r<22.5$ that were in the field and in this rate range turn 
out to be the two brightest known jovians in the field: Hermippe and Erinome (see Table~\ref{tab:characdet}).
We believe that if main belt asteroids were generating any significant confusion into
the rate cut, it is highly likely some of them would have been bright, given how shallow
the main-belt luminosity function is at these magnitudes.

Jup LIX (magnitude $\sim$23.2) had a two-year arc from observations in 2016 and 2017,
and was sufficiently close to its predicted position that identification was trivial.

Two more objects in the jovian rate range turned out to be Jup LI and LII, which is 
fortunate because \citet{Alexandersen2012} discovered them in this very same data set.
We used no knowledge of the prior existence of any irregular in our search; we have
only later identified detections in our rate range with previously known moons.
With a better photometric calibration now available, we believe the original mags of JLI and JLII were first reported as $\simeq$0.5 mag brighter than they should be.

Jup LXIX was a less trivial identification.  The 2017 and 2018 discovery astrometry
was confined to fewer dark runs than LIX and in fact its {\it predicted} sky position was just off this 2010 data set's coverage.  
However, we identified j21r113a26 as close in magnitude
and sky rate while being only about 4.1' away and thus believed this was 
a recovery of the moon.  We determined this moon should then be on archival recovery observations reported
in \citet{Alexandersen2012} that were targeted to track Jup LI and LII; we 
located and measured it.
The Minor Planet Center confirmed that these observations all link together, 
and we have thus increased the observed arc from 1 year to 8.

The object j22r94a24 was located on Palomar 5-meter observations taken on 
the same calendar date which overlap this CFHT's field's coverage; these
observations start roughly two hours before the CFHT data and overlap in
time.   

For completeness, we mention that the high-precision orbits of Harpalyke and Eurydome placed them inside this square degree,
but they occupied CCD gaps at the time of our observations and were thus not detected.
Note that this gap coverage is corrected for below in our estimates of the total
population.  Our field contained 9 known moons with roughly 20\% of the area lost
in the shift and stack process; having two non-detected moons is thus 
unsurprising.  
Lastly, there are two jovians with temporary designations
(S/2003 J 9 and S/2003 J 12)
whose ephemeris, based on a few-month arc from 2003, is on the 
field but with enormous error (of order thousands of arcseconds according to
analysis of \citet{Jacobson}).
It is possible that these moons are actually among our detections, but we
have not yet been able to establish a linkage.


To summarize, every object we detected in our jovian irregular rate cut that could get additional observations has turned out to be a jovian irregular, leaving little credence for the argument that the fainter detections in this rate range are not jovians.
While there remains a small possibility that one of these detections happens to be a small Centaur 
passing close to Jupiter, our main goal is a luminosity-function analysis to estimate the size
distribution and the total jovian irregular population down to magnitude 25.7; an interloper or 
two will have a negligible effect on these analyses.

\subsection{Slope of debiased luminosity function}

\begin{figure}[ht]
\plotone{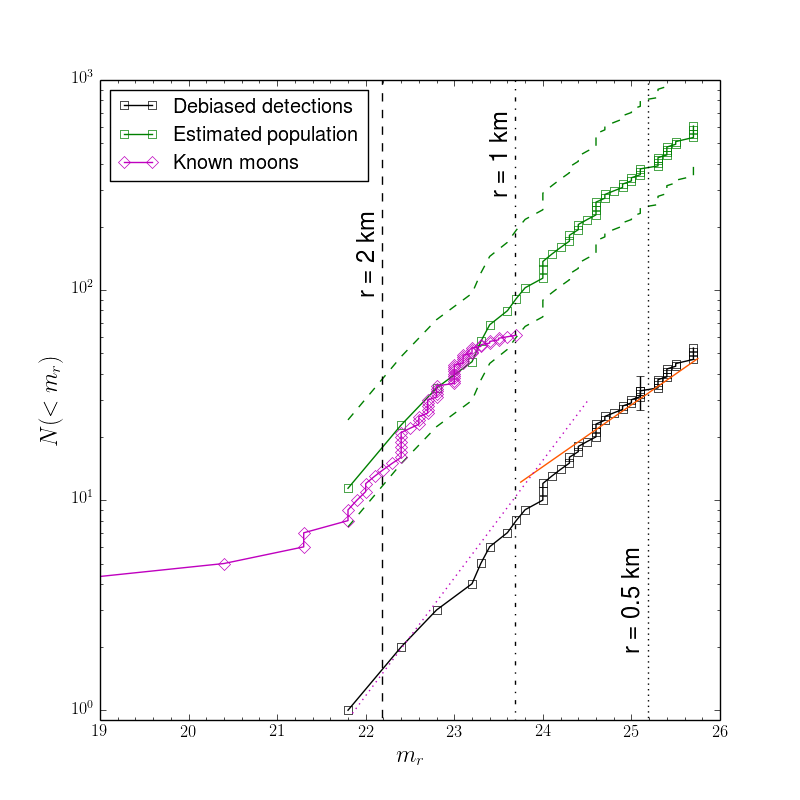}
\caption{The black line shows the cumulative debiased number of jovians based on our search field.
The magnitude corresponding to radii of 2, 1 and 0.5~km (using a 0.04 albedo and Jupiter's distance at time of the observations) are indicated by vertical lines.
The solid orange line is the $\alpha=0.29$ slope of the best-fit {\it differential} luminosity function 
(see text) that uses the debiased detections from $m_r=$23.75 -- 25.75.
We produce a total retrograde jovian population estimate (green line) and uncertainties (green dashed lines) by applying a multiplier of $11 \pm 5$ (see text) to the debiased detections; this estimate lines up nicely with the currently-known population of bright retrograde jovian irregulars (magenta line), indicating the inventory is essentially complete to $m_r\simeq23.2$.
Our differential fit to the known retrograde moons (from $m_r = 21.85$ to $23.05$, assuming no
incompleteness correction) 
has been shifted down to our debiased detections for reference (magenta dotted line), 
showing that at the bright end our detections have a similar slope but the faint end appears 
to be shallower.
Our detected sample suggests there are $\simeq$600 $m_r < 25.7$ retrograde jovians irregulars (to a factor of 2). 
}
\label{fig:lumfunc}
\end{figure}

We performed a single exponential least-squares fit on the binned differential luminosity function of both our debiased detections, from $m_r = 23.75$ to $25.75$, and on the known retrograde moons in the MPC database, starting where the size distribution starts to ramp up, $m_r =21.85$, to just before it rolls over, $23.05$ (the reason for using just the retrograde population explained in Sec \ref{sec:Pop}). 
The resulting logarithmic slopes we get from the fit for our detections and the known retrograde moons are $\alpha = 0.29 \pm 0.15$ (orange line in Fig. \ref{fig:lumfunc}) and $0.6 \pm 0.3$ respectively. 
The uncertainties were obtained by generating differential luminosity functions with random numbers for each bin drawn from a Poisson distribution with the original number in each bin being the expected number of occurrences. 
Single exponential least-squares fits were performed on 10,000 such luminosity functions, producing a distribution of slopes; we used the full width at half maximum of this distribution as our uncertainty bounds. 
The two slopes are different at a level between 1 and 2 sigma, providing weak evidence that the luminosity function for retrograde jovians changes to a shallower slope at $m_r \sim 23.5$. 
This is not a surprise, as simulations \citep{Bottke} produce `waves' in the luminosity function as the exponential index fluctuates around the collisional equilibrium value of $\alpha = 0.5$ \citep{Dohnanyi} as one descends the size distribution. 
These waves reflect the propagation down the size distribution as the largest objects in the finite distribution disrupt; the fact
that $\alpha$ is {\it close} to 0.5 usually indicates that this portion of the size distribution has `relaxed' to near 
equilibrium.

\subsection{Retrograde population estimate}
\label{sec:Pop}

We assume at this point that the vast majority of our detections have retrograde orbits. This assumption is based on 1) the edge of the field that is closest to Jupiter is about a degree away, which is approximately the projected apocentre distance of most (8 of 10) of the known direct irregular moons, and 2) only a small fraction of all known moons are direct (10/71). As such, we will make the approximation that all moons in our field are retrograde. Our analysis will thus focus on providing an estimate Jupiter's retrograde population.


We can get an estimate on the total number of jovian irregulars if we know the fraction of the population in our field's jovian offset at any time. We counted the average number of known retrograde moons in a field with the same size and with the same on-sky offset from Jupiter for 10 different oppositions (2009 to 2019). 
Note that over this 10-year time interval the moons complete many orbits and thus `lose memory' of where they were discovered (most were discovered before 2004).
On average there were 6.9 known moons in this field. Accounting for the $20\%$ of sky area lost due trimming and chip gaps, on average we will detect 5.5 known retrograde moons (of 61). Thus the multiplier going from our sample to the full populations is $11 \pm 5$; the uncertainty value comes from the standard deviation of number of known moons over the 10 oppositions.   
If there were biases induced in the fraction of known retrograde moons in our field
due to where the detections surveys found many orbits ago (which we think unlikely), there
could be a systematic present but we believe is likely small compared to our factor of
two estimate.

Our estimated total population of retrograde moons overlaps nicely with the known population from $m_r \approx 22$ to $23$ (see Fig. \ref{fig:lumfunc}). Beyond $m_r \approx 23.2$ the known population flattens out and diverges from our estimated population, indicating that the current completion limit is $m_r \approx 23.2$. This is in agreement with the fact that the brightest object which we were unable to match to a known object has $m_r = 23.3$. 

Our results produce an estimate of $160 \pm 60$ retrograde jovian moons with $m_r < 24$, which agrees with \citet{SheppardnJewitt} prediction of 100 down to the same limit (although their prediction includes retrograde and direct moons). We estimate there are $600$ retrograde jovian irregulars (within a factor of 2) down to $25.7^{th}$ magnitude. Using an albedo of 0.04 (the average retrograde jovian albedo from \citet{Grav2015}) and the distance of Jupiter at the time of observation produces a radius of $0.4$~km for a magnitude of $25.7$. We get $10\%$ error in the radius, which is dominated by the fact we don't know if the moon is half a Hill sphere in front or behind Jupiter. If retrograde jovians are almost complete down to $m_r = 23$ then the known retrograde population being well within the error bars of our population estimate suggests that we have over estimated the size our error bar; our $R>0.4$~km estimate's uncertainty may thus be less than a factor of two.

We posit that the overall completion limit for known direct jovians would likely be at a brighter magnitude that for the retrogrades. Direct jovian moons have smaller semimajor axes compared to retrograde moons, resulting in direct moons spending more time close to Jupiter. We speculate that this makes direct moons harder to detect and could contribute to the relative lack of direct jovians compared to retrograde.

\section{Conclusion}

We found 52 jovian moon candidates, seven of which were previously known, from an archival data set dating back to 2010. Using artificially implant objects we were able to turn 47 of the candidates into a luminosity function of retrograde jovian moons down to $m_r =25.7$. The slope we obtain from our luminosity function in the range of $m_r = 23.75$ to $25.75$ is $\alpha = 0.29 \pm 0.15$, which shows a weak signal of being shallower than the $0.6 \pm 0.3$ for known retrograde jovian moons in the range $m_r =21.85$ to $23.05$. Using the mean fraction of known retrograde jovian moons in a same sized field and the same offset from Jupiter as ours over 10 different oppositions, we scale our detections up to get an estimate on the total population of retrograde jovians. From our analysis we get that the current completion limit of retrograde jovian irregulars is $m_r \approx 23.2$ and there are (within a factor of two) $600$ of these moon down to $m_r = 25.7^{th}$.

\section{Acknowledgements}
This work was supported by funding from the National Sciences and Engineering Research Council of Canada.

\appendix


Table \ref{tab:astrom} in the electronic appendix provides our Minor Planet Center (MPC) format astrometry and photometry. Most faint objects have just two astrometric measurements that were obtained by stacking all and then the last half of the images. We were able to get more measurements of brighter objects, which can be seen on small subset of stacked images or even single images. Some astrometric measurements don't have photometry associated with them, either due to inability to do photometry or that we thought there were already enough reliable photometric measurements for that object.

\begin{table}[ht]
\renewcommand{\thetable}{\arabic{table}}
\centering
\caption{The format of the file containing the astrometry of both our characterised and uncharacterised detections, consistent with MPC astrometry submission. This file is available with the electronic version of this paper or at request from the authors. \label{tab:astrom} }
\begin{tabular}{cl}
\tablewidth{0pt}
\hline
\hline
\colhead{Column} & \colhead{Use}\\
\hline
    1       &  Blank \\
    2-12    &  Designation of object \\
   14       &  Blank \\
   15       &  'C' for CCD \\
   16 - 32  &  Date of middle of exposure \\
   33 - 44  &  Observed RA (J2000.0) \\
   45 - 56  &  Observed Decl. (J2000.0) \\
   57 - 65  &  Blank \\
   66 - 71  &  Observed magnitude and band \\
   72 - 77  &  Blank \\
   78 - 80  &  Observatory code \\
\hline
\end{tabular}
\end{table}



\end{document}